\documentclass[10pt,twocolumn,twoside,aps,pra,nopacs,superscriptaddress,nofootinbib]{revtex4-1}
\usepackage{graphicx}
\usepackage{amsmath,amssymb}
\usepackage{color}
\usepackage{bbm,subfigure}
\usepackage[colorlinks=false,urlcolor=blue]{hyperref}
\usepackage{theorem}
\usepackage{latexsym}
\usepackage{soul}

\newlength{\blank}
\mathchardef\ordinarycolon\mathcode`\:
\mathcode`\:=\string"8000
\def\vcentcolon{\mathrel{\mathop\ordinarycolon}}
\begingroup \catcode`\:=\active
  \lowercase{\endgroup
  \let :\vcentcolon
  }

\newcommand{\nc}{\newcommand}
\nc{\rnc}{\renewcommand}
\nc{\beq}{\begin{equation}}
\nc{\eeq}{{\end{equation}}}
\nc{\beqa}{\begin{eqnarray}}
\nc{\eeqa}{\end{eqnarray}}
\nc{\lbar}[1]{\overline{#1}}
\nc{\ketbra}[2]{|#1\rangle\!\langle#2|}
\nc{\proj}[1]{| #1\rangle\!\langle #1 |}
\nc{\avg}[1]{\langle#1\rangle}
\nc{\Rank}{\operatorname{Rank}}
\nc{\smfrac}[2]{\mbox{$\frac{#1}{#2}$}}
\nc{\tr}{\operatorname{Tr}}
\nc{\ox}{\otimes}
\nc{\dg}{\dagger}
\nc{\dn}{\downarrow}
\nc{\cA}{\mathcal{A}}
\nc{\cB}{\mathcal{B}}
\nc{\cC}{\mathcal{C}}
\nc{\cD}{\mathcal{D}}
\nc{\cE}{\mathcal{E}}
\nc{\cF}{\mathcal{F}}
\nc{\cG}{\mathcal{G}}
\nc{\cH}{\mathcal{H}}
\nc{\cI}{\mathcal{I}}
\nc{\cJ}{\mathcal{J}}
\nc{\cK}{\mathcal{K}}
\nc{\cL}{\mathcal{L}}
\nc{\cM}{\mathcal{M}}
\nc{\cN}{\mathcal{N}}
\nc{\cO}{\mathcal{O}}
\nc{\cP}{\mathcal{P}}
\nc{\cR}{\mathcal{R}}
\nc{\cS}{\mathcal{S}}
\nc{\cT}{\mathcal{T}}
\nc{\cX}{\mathcal{X}}
\nc{\cZ}{\mathcal{Z}}
\nc{\csupp}{{\operatorname{csupp}}}
\nc{\qsupp}{{\operatorname{qsupp}}}
\nc{\var}{\operatorname{var}}
\nc{\rar}{\rightarrow}
\nc{\lrar}{\longrightarrow}
\nc{\polylog}{\operatorname{polylog}}
\nc{\1}{{\openone}}
\nc{\id}{{\operatorname{id}}}

\nc{\RR}{{{\mathbb R}}}
\nc{\CC}{{{\mathbb C}}}
\nc{\FF}{{{\mathbb F}}}
\nc{\NN}{{{\mathbb N}}}
\nc{\ZZ}{{{\mathbb Z}}}
\nc{\PP}{{{\mathbb P}}}
\nc{\QQ}{{{\mathbb Q}}}
\nc{\UU}{{{\mathbb U}}}
\nc{\EE}{{{\mathbb E}}}

\nc{\qed}{{$\hfill\Box$}}

\newcommand*\gnat{\vcenter{\hbox{\includegraphics[width=.7em]{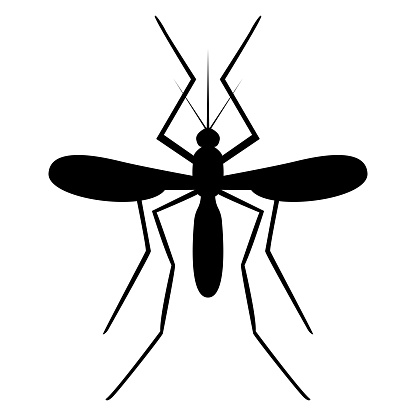}}}}
\newcommand*\gnatbig{\vcenter{\hbox{\includegraphics[width=1em]{muecke.jpg}}}}

\begin{document}

\title{A genuinely natural information measure}

\author{Andreas Winter}
%\email{andreas.winter@uab.cat}
%\affiliation{ICREA -- Instituci\'{o} Catalana de Recerca i Estudis Avan\c{c}ats, Pg.~Llu\'{\i}s Companys, 23, 08010 Barcelona, Spain}
%\affiliation{Departament de F\'{\i}sica, Grup d'Informaci\'{o} Qu\`{a}ntica, Universitat Aut\`{o}noma de Barcelona, 08193 Bellaterra (Barcelona), Spain}
%\affiliation{\st{University of Alabama Birmingham (UAB)}}
\affiliation{GIQ UAB}
%\affiliation{Universitat Aut\`onoma de Barcelona (UAB)}
%\affiliation{Working from home (WTF)}
%\affiliation{Or in the cafe, if they have wifi}
%\affiliation{Fifth affiliation}
\affiliation{ICREA}

\date{1 April 2021}

\begin{abstract}
The theoretical measuring of information was famously initiated by 
Shannon in his mathematical theory of communication, in which he proposed 
a now widely used quantity, the entropy, measured in \emph{bits}. 
Yet, in the same paper, Shannon also chose to measure the information 
in continuous systems in \emph{nats}, which differ from bits by the use 
of the natural rather than the binary logarithm. 

We point out that there is nothing natural about the choice of logarithm 
basis, rather it is arbitrary. We remedy this problematic state of 
affairs by proposing a genuinely natural measure of information, 
which we dub \emph{gnats}. We show that gnats have many advantages 
in information theory, and propose to adopt the underlying methodology 
throughout science, arts and everyday life. 
%, which unlike previous approaches unites naturalness
%with sustainability, as it is based on the virtual omnipresence of 
%specimen of the order diptera. 
\end{abstract}

\maketitle

%%%%%%%%%%%%%%%%%%%%%%%%%%%%%%%%%%%%%%%%%%%%%%%%%%%%%%%%%%%%%%%%%%%

%As science advances from ignorance to knowledge, once in a while it becomes 
%necessary to reexamine the foundations on which we are building our progress. 
%Often then we seem to regress into ignorance, as we are sometimes 
%forced by this critical activity to abandon certain basic, long-held 
%conceptions. This, however, is an illusion created in part by the pain of 
%having to let go of cherished ideas.
%Indeed, rather than impeding the progress of science, it is necessary 
%to prune false or distorted ideas from the tree of knowledge that are 
%in danger of becoming dogma. 

\textbf{Nuts and bolts of information theory.}
The ongoing digital revolution is fond of considering Claude Shannon as its 
godfather, his information theory \cite{Shannon:IT} providing 
the theoretical underpinning for the pervasive digitization of every 
kind of data, which today makes the representation and indeed 
measurement of information in \emph{bits} seem almost inevitable, if 
not entirely natural. Indeed, in the same founding paper, Shannon 
started the habit of measuring the information, and in particular 
the capacity, of continuous-variable systems in \emph{nats}. This 
parallel use of two conflicting conventions has persisted in 
information theory, and has been embedded into the grain of its 
various sub-disciplines. 

Today, after more than seventy years of information theory, which saw 
it become one of the most successful tools of all engineering branches 
occupied with the development of information processing and communication 
devices, we can state that this was an unfortunate move, as it 
imported an annoying ambiguity into the very foundations of this 
otherwise awe-inspiring theoretical framework. 
The detractors who wish to deny the severity of this well-known problem 
will inevitably point to the mathematical fact that bits and nats are 
distinguished by the choice of binary and natural logarithm, respectively, 
when defining the entropy. Indeed, for the simplest case of a uniform 
distribution on $N$ objects, 
\begin{align}
  \label{eq:log}
  S &= \log N \text{ [bits]}, \\
  \label{eq:ln}
  S &= \ln N \text{ [nats]}, 
\end{align}
are the two commonly employed formulas. They differ by a factor of $\ln 2$, 
which is well-known to be impossible to apply correctly (does it divide 
or multiply?). 

In addition, it is often claimed that the difference between 
Eqs.~\eqref{eq:log} and \eqref{eq:ln} is merely down to a choice of 
units (of information, presumably). But that is evidently nonsense, as 
entropy is, by definition, a unitless quantity. In other words, writing 
the entropy as a number, one cannot intrinsically deduce this elusive 
``unit'' from it, nor is there any way of attaching a unit in the usual 
sense that would make it unambiguous. The usual way out is to attach 
the words ``bits'' or ``nats'' 
%as we have done in Eqs.~\eqref{eq:log} and \eqref{eq:ln}, 
to indicate the calculation by which the result was obtained. 

We are thus faced with the truly fundamental problem of information science,
which has been left unattended from the beginning, but cannot be ignored 
any longer: to devise a universal measure of information, avoiding
the ultimately arbitrary choice of the basis of a logarithm. Obviously, 
despite all its success in information technology, the binary logarithm 
(basis $2$) cannot claim any more fundamental significance than the common 
one (basis $10$); as for the logarithm to basis $e$, it may be called 
``natural'', but it is not any more natural then the previous two when 
it comes to calculating entropies. 

In the present work, we finally address this issue at the very foundations 
of the field. Our approach shall be a natural one, looking for guidance in 
the physical world. While the conclusion is that henceforth we ought to 
abandon formulas like Eqs.~\eqref{eq:log} and \eqref{eq:ln}, it is hoped 
that this slight discomfort (surely overcome quickly with habit) 
will provide a sounder basis for information science.

\bigskip\noindent
\textbf{From bits to gnats.}
Our quest is to find a \emph{genuinely natural information measure}, 
which directly suggests to consider \emph{gnats}, to obtain a reasonable 
acronym. 

\begin{figure}[ht]
  \includegraphics[width=0.6\columnwidth]{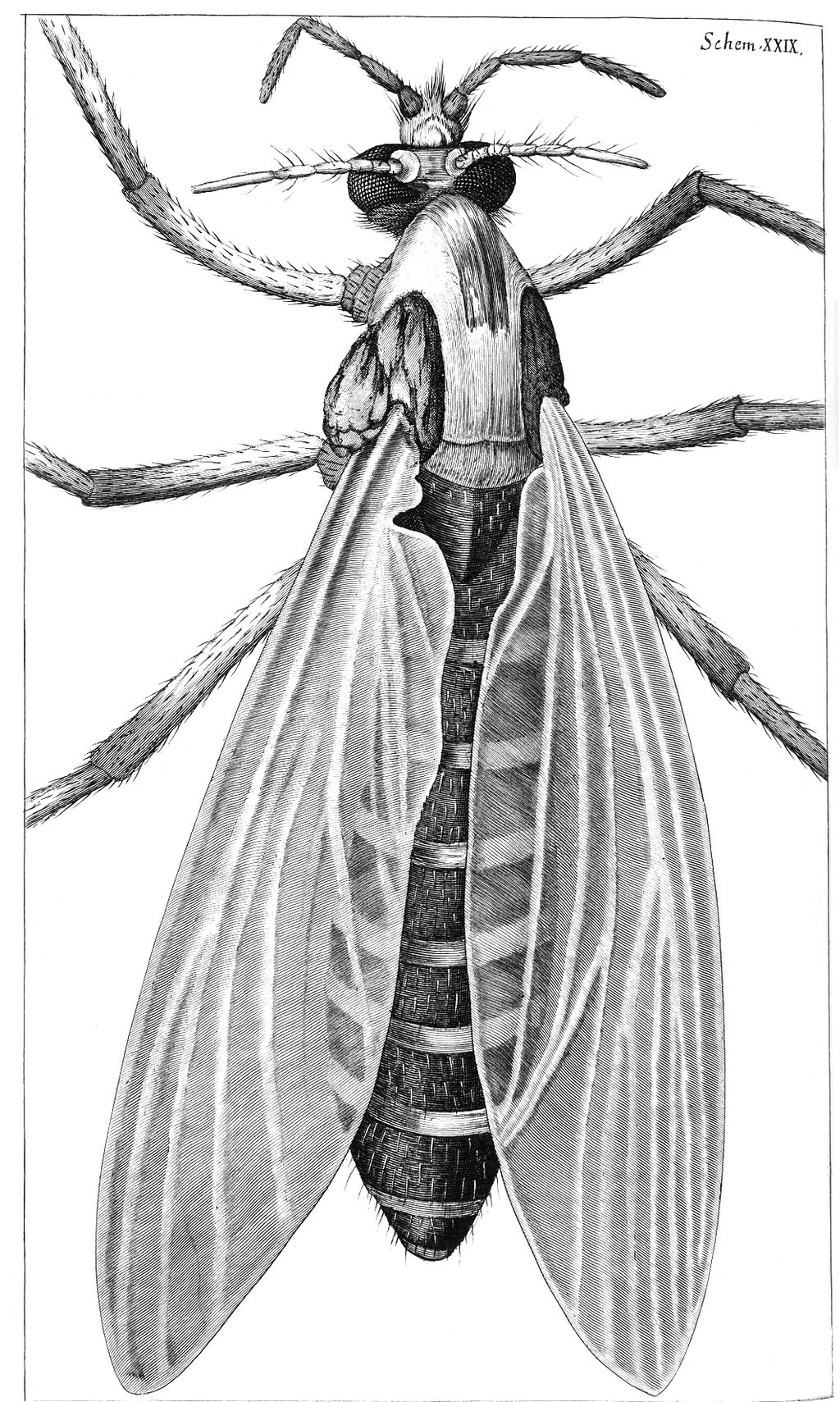}
  \caption{A gnat (after Robert Hooke's \emph{Micrographia}).}
  \label{fig:gnat}
\end{figure}

While gnats have been studied in biology for centuries
(see Fig. \ref{fig:gnat}), they have never been considered in connection 
with information technology. Indeed, the only role insects have played 
hitherto in computer science was restricted to \emph{bugs}, which is 
rather unspecific on the one hand, and on the other hand seems to 
originally refer to a completely different order (\emph{lepidoptera} 
rather than \emph{diptera}). 

However, once the idea is presented, the advantages of gnats in information 
technology are blindingly obvious: 
\begin{itemize}
  \item gnats are everywhere, and occur in unlimited supply, thus guaranteeing 
    global connectivity and sustainability; 
  
  \item because of their omnipresence, both data storage and calibration are 
    easily solved problems;

  \item gnats are discrete by design, yet entirely natural; 

  \item because of their eponymous attention span, information systems 
    based on gnats will to good approximation be memoryless, thus rendering 
    obsolete much of one-shot information theory, and restoring the wonderful 
    world of iid. 
\end{itemize}

\bigskip\noindent
\textbf{Methodology.}
Unfortunately we have to have this section, but that does not mean we 
have to waste any time on it. 

In the few words necessary, our method of discovery is 
morphological, 
lexical, and 
strictly asemantic. In these regards it follows the paradigm (a dozen indeed)
of Shannon's approach to information \cite{Shannon:IT}.
It is however influenced by certain aspects of gnomonology and 
numerology, but the reader does not have to be an expert in the latter
to understand the derivation or the result. 
Let it be enough to point out that the number seven, and hence the 
seventh letter of the Latin (English) alphabet, is of fundamental 
importance.

\bigskip\noindent
\textbf{Calibration and conversion.}
As gnats occur naturally, their comparison with bits (and for that matter, 
nats) is a question of empirical observation, see Fig. \ref{fig:gnat}. 
Initial rough measurements indicate that $1$ gnat equals $956828.3$ bits 
($663222.8$ nats); 
these values are obtained from the entropy of a simple black hole formed by an 
average gnat. The identification is justified by the fact that throwing 
it into a black hole is the only reliable way known of definitely getting 
rid of a gnat. 
As is well-known, the black hole entropy depends quadratically 
on the mass, 
\begin{equation}
  1 \text{ gnat} \equiv S_{\text{BH}}(\gnatbig) = \frac{4\pi G M_{\gnat}^2}{\hbar c},
\end{equation}
and so high-precision measurements of the gnat mass $M_{\gnat}$ will 
be vital for the future development of natural information theory. 

For the moment, a good rule of thumb is that $1$ gnat is roughly $10^6$ bits, 
meaning that we can in practice continue to use bits in information theory, 
keeping in mind that $1$ bit is roughly $1$ $\mu$gnat (micro-gnat). 
Evidently, in due course bits and nats will be \emph{defined} in terms
of gnats, as soon as a sufficiently reliable standard for the latter 
has been established.

One might ask why we do not use the Planck mass $m_P = \sqrt{\frac{\hbar c}{G}}$, 
which is only two orders of magnitude smaller than $M_{\gnat}$, 
and is supposedly a universal unit of mass. 
Well, nice try; but it won't work.

\begin{figure}[ht]
  \includegraphics[width=0.6\columnwidth]{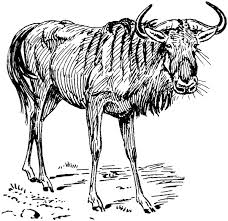}
  \caption{Holy gnu.}
  \label{fig:gnu}
\end{figure}

\bigskip\noindent
\textbf{Discussion.}
Our proposal to base information theory on gnats has been shown to unite 
several advantages: 
gnats appear naturally in discrete units, 
they are fungible, being for all practical purposes infinite in number, 
and omnipresent, meaning that global calibration is easy. 
This gives a wholly new foundation (see Fig. \ref{fig:gnu}) to information 
theory, which we expect to have profound repercussions.

\begin{figure}[ht]
  \includegraphics[width=0.7\columnwidth]{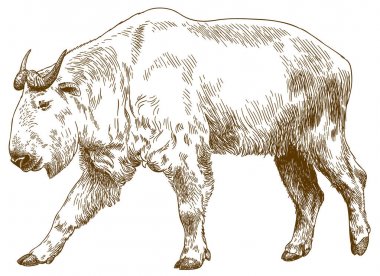}
  \caption{A pretty gnu goat.}
  \label{fig:gnugoat}
\end{figure}

Our novel approach points to a more fundamental principle, 
which has the potential to revolutionise all gnatural  
sciences, but also medicine \cite{Pompey}, 
psychology \cite{Pompey-2}, 
and economics \cite{Gnash}, 
all the way to literature \cite{GregorSamsa} 
and psychological self-help \cite{Gnome}.
The possibilities are endless \cite{Shannon:bandwagon}
(see Fig. \ref{fig:gnugoat}).

Not all conceptual problems are solved by our proposal, and 
it leaves in particular a number of open questions, not least 
in relation to quantum physics and quantum information theory. 
First, black hole evaporation and holography are waiting to 
be reformulated in terms of the new information measure. 
In fact, we conjecture that the interior of black holes has a 
consistent description entirely in terms of gnats; consequently, 
when a black hole evaporates, it should do so by releasing all 
its gnats. 
Secondly, in quantum information theory there are other types 
of information beyond the bit, most importantly qubits and 
ebits, cf.~\cite{quantum-leap}. 
It remains an open question how to quantise gnats, since ``qnats'' 
is not very convincing. Among the proposals currently being 
evaluated are knats and knots.

\bigskip\noindent
\textbf{Acknowledgments.}
The author thanks an anonymous gnarled gnof for sharing their gnaphalium and 
a priceless gnetum gnemon during the preparation of the present article. 
Gnawing at gnocchi, knocking them back knowingly, and persistent gnashing 
and gnarring were essentially supported by assorted gnathostomata, which are 
gratefully acknowledged.

\bibliographystyle{plain}

\end{document}